\documentclass[10pt,journal,compsoc]{IEEEtran}
\usepackage{graphicx}
\usepackage{epsfig}
\usepackage{subfigure}
\usepackage{multirow}
\usepackage{algorithm,algpseudocode}
\usepackage{url}
\usepackage{tikz}
\usepackage{amsmath}
\usepackage{mathtools,amssymb,lipsum,caption}

\ifCLASSOPTIONcompsoc
  \usepackage[nocompress]{cite}
\else
  \usepackage{cite}
\fi
\ifCLASSINFOpdf
\else
\fi

\begin{document}
\title{Hierarchical System Mapping for\\Large-Scale Fault-Tolerant Quantum Computing}

\author{Yongsoo~Hwang
        and~Byung-Soo~Choi%,~\IEEEmembership{Life~Fellow,~IEEE}% <-this % stops a space
\IEEEcompsocitemizethanks{
\IEEEcompsocthanksitem 
Y. Hwang and B.-S. Choi are with Electronics and Telecommunications Research Institute, Daejeon, Republic of Korea, 34129.\protect\\
E-mail: bschoi3@etri.re.kr
}% <-this % stops an unwanted space
}

\IEEEtitleabstractindextext{%
\begin{abstract}
Considering the large-scale quantum computer, it is important to know how much quantum computational resources is necessary precisely and quickly. 
Unfortunately the previous methods so far cannot support a large-scale quantum computing practically and therefore the analysis because they usually use a non-structured code. 
To overcome this problem, we propose a fast mapping by using the hierarchical assembly code which is much more compact than the non-structured code. 
During the mapping process, the necessary modules and their interconnection can be dynamically mapped by using the communication bus at the cost of additional qubits. 
In our study, the proposed method works very fast such as 1 hour than 1500 days for Shor algorithm to factorize 512-bit integer.
Meanwhile, since the hierarchical assembly code has high degree of locality, it has shorter SWAP chains and hence it does not increase the quantum computation time than expected.
\end{abstract}

}

\maketitle

\IEEEdisplaynontitleabstractindextext

% \fi
\IEEEpeerreviewmaketitle

\IEEEraisesectionheading{\section{Introduction}\label{sec:introduction}}

\IEEEPARstart{T}{he} era of quantum computing already started.
Several gigantic IT corporations have been devoted to develop quantum computing devices, and some of them are now providing quantum computing cloud service to the public~\cite{IBM:Iidq8v80,Rigetti:t5QgvCqz}.
But the scale of the devices are still small (dozens of qubits), and therefore applications for them are also very limited. 
While we look forward to seeing a quantum supremacy with such a small scale quantum device soon, most of the public have interest in a large scale universal quantum computer that can run large-sized quantum algorithms to solve real world problems.

Suppose that you have a quantum computing hardware and a quantum algorithm. 
What do you have to do to run the algorithm with the hardware?
Quantum algorithm is a logical description of how to solve a given problem.
It is usually based on ideal quantum computing hardware with noiseless gate and long qubit interaction.
However, in reality, a quantum computer as a physical entity has a certain physical and logical limitation. 
For example, qubits might be peculiarly arranged and connected with each other (see Ref.~\cite{IBM:Iidq8v80,Rigetti:t5QgvCqz}).
Therefore, to run a quantum algorithm on a quantum computer, we have to prepare an architecture-specific description of a quantum algorithm beforehand.
This is why a quantum system mapping is necessary.

The principle of a quantum system mapping is straightforward.
Initialize a quantum computer architecture and recast a quantum algorithm for the architecture.
A detailed procedure completely depends on a quantum algorithm, a quantum computer architecture and a mapping algorithm.
An architecture-specific description of a quantum algorithm, the main product of the system mapping, is called a \textit{system code} in this work.
It is possible to run a quantum algorithm on the quantum computer by executing the system code. 
Another output of the system mapping is an \textit{expected performance} of a quantum computing.
Since the system mapping treats all quantum instructions of a quantum algorithm under the pre-assumed quantum computer architecture, it is possible to analyze some performance of a quantum computing, i.e., a circuit depth and an execution time.

Quantum algorithm for the system mapping is provided in a quantum assembly code format.
A quantum assembly code (QASM) is an intermediate representation of a quantum algorithm between an abstract description and a physical machine instruction description~\cite{Cross:2017ud,JavadiAbhari:2015jf,Svore:2006iw}.
It is a list of quantum instructions denoting a combination of a quantum gate and target qubit(s), and is generated through a compile by taking a programmed quantum algorithm.
Due to the lack of standard for QASM, the specific representations may be slightly varied~\cite{Rigetti:t5QgvCqz,Cross:2017ud,Green:2013ha,JavadiAbhari:2015jf}. %,Khammassi:2018ug}.

There are two kinds of QASMs in terms of structure: \textit{non-modular} type and \textit{modular} type.
A non-modular QASM na\"{i}vely enumerates all quantum instructions. 
On the other hand, a modular QASM is hierarchically structured with multiple functions called a \textit{module}.
A module corresponds to a composite quantum operation composed of multiple quantum instructions including calling other sub-modules. 
In general, a modular QASM is composed of one \textit{main} module and multiple sub-modules~\cite{JavadiAbhari:2015jf}.

To date most quantum system mappings have been performed based on the non-modular QASM.
This is because the mapping is straightforward due to the trivial structure of the QASM. 
Since the quantity of the developed qubits in reality is very small as mentioned above, small-sized quantum algorithms have been targeted for the system mapping and for such situations the system mapping with the non-modular QASM has been working well.

However, when we treat a large-sized quantum algorithm, practically serious problems arise with the non-modular QASM.
Please recall that a non-modular QASM is a simple list of quantum instructions.
As the size of a quantum algorithm increases, so definitely does the size of a non-modular QASM.
It completely follows along the complexity of the algorithm.
We obtained a 39 TB sized non-modular QASM for Shor algorithm to factorize a 512-bit integer (see TABLE~\ref{tab:shor_file_size}).
Due to the lack of classical storage and memory, we could not even attempt to generate a non-modular QASM for a larger quantum algorithm.
Obviously, the size of the most interested problems for a quantum computing are beyond the capacity of classical super-computing.
Therefore, we will definitely have to deal with a more larger QASM, and in doing so practical problems caused by such enormous sized QASM will be one of critical issues in classical control part of a quantum computing. 

\begin{table}[t]
\centering
\caption{
The QASM size comparison over Shor's factoring algorithm.
The QASMs are generated by using the compiler ScaffCC~\cite{ScaffCCScaffCC:vm,JavadiAbhari:2015jf}.
}
\begin{tabular}{c||c|c} \hline
Input Size & Non-Modular & Modular \\ \hline
128 & 1.7 TB & 23.5 MB \\
256 & 14.2 TB & 88.1 MB\\
512 & 39.0 TB & 338.6 MB \\ \hline
\end{tabular}
\label{tab:shor_file_size}
\end{table} 

%\begin{figure}[t]
%\centering
%\epsfig{file=Shor_file_size.eps, scale=0.22}
%\caption{
%The QASM size comparison over Shor's algorithm for factoring 128, 256, and 512 bit integers. 
%The QASMs are generated by using the compiler ScaffCC~\cite{ScaffCCScaffCC:vm,JavadiAbhari:2015jf}.
%}
%\label{fig:shor_file_size}
%\end{figure}

In this work, we present a quantum system mapping for a large-scale fault-tolerant quantum computing. 
To this end, we turn our attention to a modular QASM instead.
As mentioned above, a modular QASM is hierarchically structured as composed of modules, and we found that such hierarchical structure can suppress the scalability in the size of QASM.
Suppose that there is an $n$-qubit composite quantum operation $U$ composed of $K$ quantum gates and it is called as much as $N$ times. 
To represent such $N$ iterations, a non-modular QASM requires $K\times N$ quantum instructions, but a modular QASM only requires $K+N$ instructions by defining $U$ as a module.
Empirically, the values of $K$, $N$ are not small in a non-trivial quantum algorithm\footnote{In general, $K$ is $O(10^2)$ and $N$ is $O(10^4)\sim O(10^6)$ in Shor algorithm.}.
%Empirically, the most frequently called modules are ones that decompose an $R_Z(\theta)$ gate into a sequence of hundreds $H$, $S$ and $T$, and such modules are called as much as hundreds of thousands times in quantum algorithm\footnote{$M$ is in $O(10^2)$ and $N$ is in $O(10^4)\sim O(10^6)$. Furthermore, there are thousands of modules to decompose $R_Z(\theta)$ gates in Shor algorithm.}.
%Please note an $R_Z(\theta)$ gate for an arbitrary rotation angle $\theta$ is not implemented in the fault-tolerant manner, and therefore it should be decomposed into $H$, $S$ and $T$ gates those are fault-tolerantly implementable.

In this regards, a modular QASM is much smaller than a non-modular QASM (see Table~\ref{tab:shor_file_size}).
Therefore, it is possible to generate and manage a QASM for a large-sized quantum algorithm.
By the way, currently, only a few quantum compilers support a modular (or hierarchically structured) QASM.
This work is currently compatible with open-source quantum compiler ScaffCC~\cite{ScaffCCScaffCC:vm,JavadiAbhari:2015jf}.
In this work, we describe how to exploit the modular structure of the modular QASM on a pre-assumed quantum computer, and discuss the strengths and weaknesses of the proposed mapping.

\section{Hierarchical System Mapping}\label{sec:mapping_modular}
\subsection{Hierarchically Structured Qubit Layouts}\label{subsec:qubit_layout}

\begin{figure}[t]
\centering
\epsfig{file=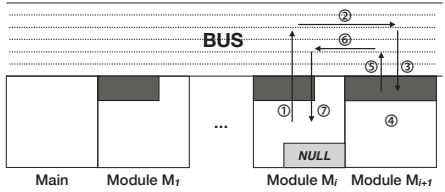, scale=0.23}
\caption{
The mapping process for calling a module is composed of seven steps: 1. (forward) move qubits to the bus, 2. (forward) move to the target module, 3. (forward) move to the parameter qubit cells (dark grey cells), 4. module operations, 5. (backward) move qubits to the bus, 6. (backward) move to the original module, and 7. (backward) move to the original qubit positions.}
\label{fig:module_call}
\end{figure}

System mapping begins with the initialization of a quantum computer architecture, i.e., a qubit array.
A qubit array is not specifically limited, but in this work we assume that it is hierarchically structured. 
A quantum computer is then made of modules (computing regions) and a communication bus connecting them.
Modules and bus are composed of logical qubits encoded by quantum error-correcting codes.
At a module, qubits are manipulated by following QASM, and transmitted between modules through the communication bus.
The bus makes a quantum computing more reliable because a logical data qubit is only interacted with ancilla qubits on the bus and therefore a quantum error does not propagate to other data qubits.
Furthermore, the communication bus permits parallel qubit movements and therefore, as will be discussed later, the hierarchically structured quantum computer can make a quantum computing more efficient. 
Fig.~\ref{fig:module_call} shows an example of a quantum computer architecture where modules are arranged on the 1D  array and qubits in each module are arranged on the 2D array.

%We need to discuss qubits in a modular QASM in detail.
In a modular QASM, qubits are classified into two types: \textit{local} qubits and \textit{parameter} qubits.
Local qubits are initialized, manipulated and measured within a module, whereas parameter qubits are passed between modules over a communication bus.
Therefore, a qubit array requires physical space for both qubits.
In Fig.~\ref{fig:module_call}, the dark grey cells indicate parameter qubits, and white cells represent local qubits.
Besides, dummy qubits (or empty space), light grey cells denoted by ``NULL", are sometimes required to form the 2-dimensional rectangular shape of a module.

Qubit that resides inside a module supports universal quantum computing.
The logical qubit is composed of data qubits for holding data and ancilla qubits for quantum error correction and logical operations. %~\cite{Svore:2007tb,Fowler:2012fi}.
On the other hand, qubits composing a communication bus do not need to support universal quantum computing, in particular logical non-transversal gates.
Therefore the composition of logical qubits for computing and communication may be slightly differ according to quantum error-correcting codes.

\subsection{Mapping Algorithm}\label{subsec:mapping}
The proposed system mapping proceeds module by module starting from \textit{main} module.
For mapping a module, we first allocate a physical space for the module next to the previously allocated module, and arrange the qubits of QASM on the space.
The space is made of cells as much as the number of qubits in the module. 
Here\footnote{In Section~\ref{sec:discussion}, we will discuss an optimized qubit placement of a module.}, we simply arrange the qubits in the first-come-first-served manner.
We then read each quantum instruction from QASM, and conduct an appropriate mapping process. 
After mapping a module, the allocated space for a module is de-allocated.

The proposed mapping requires two kinds of lookup tables, one \textit{global} table and several \textit{local} tables.
These tables will be used to record the performance of modules and of qubits of each module.
On beginning the mapping of a module $M$, a local table is initialized as $(time[q_i]=0, cycle[q_i]=0)$ over $i=1\cdots n$, where $n$ is the number of qubits of the module.
Note that $time[q_i]$ ($cycle[q_i]$) is a metric to measure an execution time (a circuit depth) of a module. 
As the mapping proceeds, the metric is being updated as $time[q_i]=time[q_i]+wait_t+u_t$ ($cycle[q_i]=cycle[q_i]+wait_c+1$).
Note that $u_t$ is the execution time of a quantum gate $u$, and $wait_t$ ($wait_c$) is a required waiting time (cycle) for a multi-qubit quantum operation. 
On finishing the mapping, the performance of the module $M$ will be recorded on the global table by picking the maximum values from the local table, $time[M] = \max_i \{ time[q_i] \}$ and $cycle[M] = \max_i \{cycle[q_i] \}$.

In a modular QASM, there are three kinds of quantum instructions: \textit{1}- and \textit{2}-qubit gate and \textit{module}.
The mapping of the 1-qubit quantum gate is straightforward.
It can be done independently from other qubits by arranging the gate to a target qubit at the proper time $time[q_i]$. 
The execution of the gate finishes at time $time[q_i]+u_t$. 
On the other hand, the mapping of a 2-qubit gate requires that two qubits have to be ready spatially and temporally.
To the best of our knowledge, the feasible implementation of a 2-qubit gate is local.
If two qubits are located at distant, they should be placed adjacently by moving qubits.
Besides, the gate can act when both qubits are in idle status in common, $\max \{time[q_i], time[q_j] \}$, where $time[q_i]$ ($time[q_j]$) is the time previous operation on the qubit $q_i$ ($q_j$) is finished.
%Besides, the gate can act when previously scheduled operations for both qubits are finished, $\max \{time[q_i], time[q_j] \}$, where $time[q_i]$ ($time[q_j]$) is the time previous operation on the qubit $q_i$ ($q_j$) is finished.

% module
The third type quantum instruction, a module, seems like a multi-qubit quantum function.
Therefore, on the surface, the required mapping process for a module is similar with that for a 2-qubit gate.
The execution of a module begins when all the qubits are temporally and spatially in ready. 
The only difference from the case of a 2-qubit gate is, as mentioned above, that a physical space is allocated for a module.
Therefore, to perform the mapping of a module, we have to consider qubit movements between a present (calling) module and a target (called) module. 
%After the operations at the target module, the transmitted qubits have to be back to their original locations.

Suppose that a module $M_i$ is being mapped now, and we have to process an instruction ``$M_k (q_a, q_b, q_c)$", calling a module $M_k$ with qubits $q_a$, $q_b$ and $q_c$. 
For that, we pause the mapping of the module $M_i$ and turn our attention to the mapping of the module $M_k$.
We allocate a physical space for the module $M_k$ next to $M_i$, and pass the qubits to the parameter qubit area of $M_k$ through a communication bus.
We call this movement a \textit{forward} qubit passing.
After all qubits are placed for their designated locations, quantum instructions of $M_k$ are performed.
If another module is called again before the finish, the mapping of the present module is paused and the above-mentioned processes for the newly called module are performed recursively.
After executing all instructions of $M_k$, the passed qubits have to be back to the original place of $M_i$.
This is called a \textit{backward} qubit passing.
After the mapping of $M_k$, we de-allocate the space for $M_k$ and continue the mapping of $M_i$.
Note that to trace the mappings of modules, a stack\footnote{A typical data structure for first-in-last-out in computer science} can be very useful.

We have mentioned that a module plays as a composite quantum operation, and is called many times during the execution of a quantum algorithm.
Whenever a module is called, it always works in the same way for the argument qubits placed in the same location.
Which means that by mapping a module once, it is possible to get a system code and a performance of a module completely.
Note that we have to measure the distance for the qubit passings every time when calling a module.
In this regards, during the mapping if we are faced with a module that has never been mapped before, we perform the mapping for the called module.
Otherwise, we just refer the mapping result of the module recorded in the global lookup table instead.

By performing the mapping for a module only once, it is possible to reduce the duration of the mapping non-trivially than the non-modular mapping.
On considering the above case that $K$ quantum instructions are repeated $N$ times, the non-modular mapping requires the mapping as much as the number of such iterations.
As mentioned before, the numbers of the repetitions $N$ and the bunches of quantum instructions (corresponding to a module) are so large for a non-trivial quantum algorithm.

%While the non-modular mapping requires the mapping as much as the number of such iterations, in the proposed mapping only one time mapping is enough regardless of the number of iterations.
%
%Suppose that we are mapping a module $M_i$ now, and a next quantum instruction is calling a module $M_j$.
%If the module $M_j$ has never been mapped before, we pause the mapping of the module $M_i$ and start the mapping of the module $M_j$.
%If the module $M_j$ includes calling an unmapped module $M_k$, we also pause the mapping of $M_j$ and do the mapping of $M_k$.
%After the mapping the module $M_j$, we continue the mapping of the module $M_i$.
%Later, the module $M_j$ is called again, then we do not perform the mapping of the module again, instead we just refer the previous mapping result recorded in a global lookup table.
%Note that to trace the mappings of modules, a stack\footnote{A typical data structure for first-in-last-out in computer science} can be very exploited.

\section{Discussion}\label{sec:discussion}

We discuss the strengths of the proposed mapping against the non-modular mapping.
First, it is possible to perform the mapping for a large-sized quantum algorithm.
As mentioned above, for a same quantum algorithm, a modular QASM is much smaller than a non-modular QASM. 
Therefore, it becomes possible to generate and manage QASM for a larger quantum algorithm (see TABLE~\ref{tab:shor_file_size}).
%Practically, ScaffCC compiler requires less time to generate a modular QASM~\cite{JavadiAbhari:2015jf}.

\begin{figure}[t]
\centering
\epsfig{file=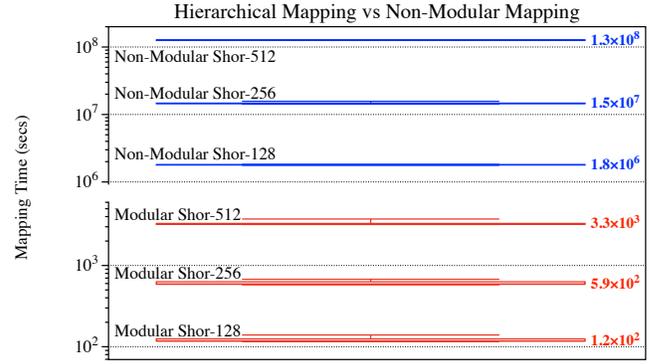, scale=0.23}
\caption{
The comparison in the required time between the hierarchical mapping (red) and the non-modular mapping (blue).
The denoted numerical value indicates mean time from several repetition. 
}
\label{fig:mapping_time}
\end{figure}

Second, it takes much smaller time to perform the mapping. 
This is due to the less-sized QASM and the one-time module mapping.
Fig.~\ref{fig:mapping_time} compares the required mapping time between both QASMs.
We have implemented both mappings with Python, and tested them on a computer system of 3.5 GHz CPU and 128 GB Memory.
The required time for the non-modular mapping are estimated from the statistics obtained from the proposed mapping.
The mapping for Shor-512 with a non-modular QASM requires 1500 days, but the proposed mapping can be done within 1 hour.

%\begin{table}[t]
%\centering
%\caption{
%The comparison in the required time between the hierarchical mapping and the non-modular mapping.
%}
%\begin{tabular}{c||c|c} \hline
%\multirow{2}{*}{Algorithm} & \multicolumn{2}{c}{Mapping Time} \\ \cline{2-3}
%					& Non-Modular & Hierarchical \\ \hline
%Shor-128 & $2.022477 \times 10^{10}$ & $2.022470 \times 10^{10}$\\
%Shor-256 & $1.582834 \times 10^{11}$ & $1.582831 \times 10^{11}$\\
%Shor-512 & $1.319982 \times 10^{12}$ & $1.319981 \times 10^{12}$ \\ \hline
%\end{tabular}
%\label{tab:optimization}
%\end{table} 

Third, it is possible to analyze and optimize a physically implemented quantum algorithm efficiently.
Through the proposed mapping, we can find out which modules are bottlenecks of the algorithm and/or the most frequently executed.
Then, by improving such modules, we can make the algorithm better.
At the same time, a pre-analysis of QASM required for an optimization can be done very efficiently in the proposed mapping. %unlike the standard mapping.
As an example, we optimize a qubit placement~\cite{Shafaei:2014fl,Siraichi:2018bg,Pedram:2016gh} for a module.
By doing so, the quantity of SWAP operations required for a CNOT gate between distant qubits can be reduced.
In this work, we apply linear programming approach~\cite{Shafaei:2014fl} for Shor and BWT (Binary Welded Tree)~\cite{JavadiAbhari:2015jf,Childs:2003vb} algorithms.
By the way, unfortunately, the performance gain by the optimization in case of Shor algorithm is negligible (circuit depth reduction less than 0.01\%), but the degree of the improvement depends on a quantum algorithm. 
We observed that the circuit depth of BWT algorithm can be reduced by 1$\sim$6\%, i.e., $2.42\times 10^5 \rightarrow 2.26\times 10^5$ for BWT-10.

\begin{table}[t]
\centering
\caption{
The quantum resource between the hierarchical mapping and the non-modular mapping in Shor algorithm.
The qubits in the hierarchical mapping represent ``computing qubits (bus qubits)".
We determine the quantity of the bus qubits with the assumption that the bandwidth of the bus equals to the maximum number of parameter qubits.
}
\begin{tabular}{c||c|c|c|c} \hline
\multirow{2}{*}{Algorithm} & \multicolumn{2}{c|}{Non-Modular} & \multicolumn{2}{c}{Hierarchical} \\ \cline{2-5}
					& Qubits & SWAPs & Qubits & SWAPs\\ \hline
Shor-4 & 46 & $4.08\times 10^4$ & 115 (374) & $1.23 \times 10^6$ \\
Shor-8 & 90 & $4.71\times 10^5$ & 227 (1,122) & $1.02 \times 10^7$ \\
Shor-16 & 178 & $5.82\times 10^6$ & 448 (3,055) & $7.66\times 10^7$\\ \hline
\end{tabular}
\label{tab:quantum_resource}
\end{table}

We now discuss the weaknesses of the proposed mapping in terms of quantum resource.
The proposed mapping assumes a hierarchically structured quantum computer composed of multiple computing regions (module) and a communication bus connecting them.
Without doubt, the proposed mapping requires more qubits and qubit movements (see Table~\ref{tab:quantum_resource}).
On average, 2.5 times more computing qubits and additionally arbitrary bus qubits are necessary, and more SWAPs are performed on the increased qubits.

However, we observed that surprisingly the length of a quantum computing does not increase as much as exactly the increased SWAPs.
This is because the proportion of SWAPs to total gates is small (10\%) and most of the SWAPs (99\%) are utilized in the qubit passings where parallel SWAPs are allowed.
%The average number of the parameter qubits for the qubit passings are gradually increasing along the input size.
We observed that only a half or a third of SWAP gates affect the quantum computing circuit depth.
Besides, as shown in TABLE~\ref{tab:quantum_resource}, the difference in the quantity of SWAPs between both mappings decreases as the input size increases ($30.3$X$\rightarrow$$21.8$X$\rightarrow$$13.16$X).
%Along this line, we guess that in the large-sized quantum algorithm the quantity of SWAPs in the hierarchical mapping approaches to the quantity of SWAPs in the non-modular mapping, or can be smaller in the best case.
Along this line, we guess that in the large-sized quantum algorithm the difference in the quantity of SWAPs between both mappings will be vanishing, or can be smaller in the best case.

We also need to emphasize that the quantity of SWAPs for a CNOT within a module is negligible.
In the proposed mapping, qubits for mutual interaction are definitely positioned in the same module, and the area of each module is smaller than the space for the non-modular mapping.
For Shor algorithm ($N$=4, 8, 16), the proposed mapping only requires on average 0.04, 0.02 and 0.02 SWAPs for a CNOT gate, whereas the non-modular mapping requires 1.00, 1.42 and 2.18 SWAPs.
As the input size increases, the quantity of SWAPs for the proposed mapping stays as 0.009 (Shor-128), 0.007 (Shor-256) and 0.005 (Shor-512), but the non-modular mapping requires more SWAPs for a CNOT gate.

In this regards, we believe that, in the large-scale quantum computing regime, the proposed mapping makes a quantum computing can be done with less running time.
Furthermore, the difference in the quantity of qubits between both mappings can be reduced by controlling the area of the communication bus, in particular the bandwidth of the bus.
Reducing the bandwidth may calls for the delay of SWAPs in some rare cases.
Note that the qubit passing rarely exploits the full bandwidth of the bus.

\section{Conclusion}\label{sec:conclusion}

Our future work is to apply the most realistic quantum computer architecture~\cite{Lin:2015bv} and to improve the mapping algorithm. 
In the present work, we assume all qubits work concurrently.
However, in reality, some qubits take longer time for error correction and/or non-transversal logical gates. 
In doing so, the gate scheduling may be in trouble.
A communication bus also has to be controlled more delicately.
Currently, we assume the bus works on demand and qubit passings are done on time.
But, we believe practically network traffic may disturb such ideal communication, and the bus may be in congestion, in particular for highly parallel quantum application~\cite{JavadiAbhari:2017kx}.
Lastly, the proposed mapping currently uses many qubits (physical space), but majority of them just wait from calling other modules during the most of the execution time.
In the future work, we will improve the spatial efficiency by the mapping algorithm.

\ifCLASSOPTIONcompsoc
  \section*{Acknowledgments}
\else
  \section*{Acknowledgment}
\fi
This work was supported by Electronics and Telecommunications Research Institute (ETRI) grant funded by the Korean government [17ZH1200, Research and Development of Quantum Computing Platform and its Cost-Effectiveness Improvement].

\ifCLASSOPTIONcaptionsoff
  \newpage
\fi

\end{document}